\documentclass[journal, 10pt]{IEEEtran}
\usepackage{amsthm,amssymb,graphicx,multirow,amsmath,color,amsfonts}
\usepackage[update,prepend]{epstopdf}
\usepackage[noadjust]{cite}
\usepackage[latin1]{inputenc}
\usepackage{tikz}
\usetikzlibrary{arrows,calc} 
\usepackage{pdfpages}
\usepackage{tabulary}
\usepackage{multirow}
\usepackage{comment}
\usepackage{dsfont}
\usepackage{subfigure}
\usepackage{algorithm}
\usepackage{algpseudocode}
\algnewcommand\algorithmicinput{\textbf{Input:}}
\algnewcommand\algorithmicoutput{\textbf{Output:}}
\algnewcommand\Input{\item[\algorithmicinput]}
\algnewcommand\Output{\item[\algorithmicoutput]}

\def\nbb{{\mathbf{b}}}

\def\nbp{{\mathbf{p}}}
\def\nbq{{\mathbf{q}}}

\def\nbs{{\mathbf{s}}}

\def\nbx{{\mathbf{x}}}

\def\nbA{{\mathbf{A}}}

\def\nbP{{\mathbf{P}}}

\def\ncalA{{\mathcal{A}}}

\def\ncalC{{\mathcal{C}}}

\def\ncalS{{\mathcal{S}}}

\def\nbbE{{\mathbb{E}}}

\def\nbbR{{\mathbb{R}}}

\def\nbbV{{\mathbb{V}}}

\def\nrmd{{\rm d}}

\def\nrmm{{\rm m}}

\newtheorem{lemma}{Lemma}

\newtheorem{remark}{Remark}

\def\argmin{\operatorname{arg~min}}

\def\figref#1{Fig.\,\ref{#1}}%

\allowdisplaybreaks 
\IEEEoverridecommandlockouts 

\begin{document}
\graphicspath{{./Figures/}}

\title{
A New Statistical Approach to Calibration-Free Localization Using Unlabeled Crowdsourced Data
}

\author{
Haozhou Hu,~\IEEEmembership{Student Member,~IEEE}, Harpreet S. Dhillon,~\IEEEmembership{Fellow,~IEEE}, R. Michael Buehrer,~\IEEEmembership{Fellow,~IEEE}
\thanks{Preliminary results from this work were presented at ICC 2025~\cite{conf3}. The authors are with Wireless@VT, Bradley Department of Electrical and Computer Engineering, Virginia Tech, Blacksburg, VA, 24061, USA. Email: \{haozhouhu, hdhillon, rbuehrer\}@vt.edu. The support of the US NSF (Grants CNS-2107276 and CNS-2225511) is gratefully acknowledged. This work was also partly supported by the Commonwealth Cyber Initiative, an investment in
advancing cyber R\&D, innovation, and workforce development. For more information about
CCI, visit www.cyberinitiative.org.
} 
}

\maketitle

\begin{abstract}
Fingerprinting-based indoor localization methods typically require labor-intensive site surveys to collect signal measurements at known reference locations and frequent recalibration, which limits their scalability.
This paper addresses these challenges by presenting a novel approach for indoor localization that utilizes crowdsourced data {\em without location labels}.
We leverage the statistical information of crowdsourced data and propose a cumulative distribution function (CDF) based distance estimation method that maps received signal strength (RSS) to distances from access points.
This approach overcomes the limitations of conventional distance estimation based on the empirical path loss model by efficiently capturing the impacts of shadow fading and multipath.
Compared to fingerprinting, our {\em unsupervised} statistical approach eliminates the need for signal measurements at known reference locations.
The estimated distances are then integrated into a three-step framework to determine the target location.
The localization performance of our proposed method is evaluated using RSS data generated from ray-tracing simulations. Our results demonstrate significant improvements in localization accuracy compared to methods based on the empirical path loss model.
Furthermore, our statistical approach, which relies on unlabeled data, achieves localization accuracy comparable to that of the {\em supervised} approach, the $k$-Nearest Neighbor ($k$NN) algorithm, which requires fingerprints with location labels. For reproducibility and future research, we make the ray-tracing dataset publicly available at~\cite{croudrt}.
\end{abstract}

\begin{IEEEkeywords}
Fingerprinting, Indoor localization, Wireless sensor networks, Unsupervised learning, WiFi positioning system.
\end{IEEEkeywords}

\section{Introduction}
Over the past decade, interest in indoor location-aware applications, such as navigation, asset tracking, and personalized services in smart buildings, has grown significantly. While the Global Positioning System (GPS) provides accurate outdoor localization, its signals struggle to penetrate most building materials, making it ineffective indoors.
Indoor signal propagation differs significantly from outdoor environments due to higher attenuation from walls and furniture, increased multipath effects from reflections, and significant interference from nearby devices. 
To address these challenges, several commercial localization systems integrate multiple positioning techniques to operate seamlessly in both indoor and outdoor environments. Companies like Apple and Google use hybrid localization approaches that combine GPS, WiFi, cellular signals, and Bluetooth to determine a device's position~\cite{apple_core_location, google_fused_location}.
GPS offers broad global coverage for outdoor navigation but can be unreliable in indoor environments. Existing indoor localization solutions can generally be categorized into two approaches: those that rely on specialized hardware, such as Ultra-Wideband~\cite{9810941}, Infrared~\cite{5647415}, and Ultrasound~\cite{9906111}, which requires the extensive deployment of dedicated infrastructure for localization purposes; and those that leverage existing infrastructure, such as WiFi or Bluetooth~\cite{8692423, 8371230}, using off-the-shelf wireless hardware commonly equipped in handheld devices.
WiFi access points support rapid location estimation by associating scanned network identifiers with extensive, crowdsourced databases. Additionally, Bluetooth Low Energy (BLE) beacons, such as Apple's iBeacon, enable precise indoor positioning by measuring signal strength at close range.


WiFi-based indoor positioning systems are especially attractive due to their seamless integration with existing infrastructure and compatibility with most off-the-shelf devices, making their deployment straightforward~\cite{overview3}.
There are two major localization approaches in WiFi-based positioning systems: (1) geometric approaches, which analyze the characteristics of received signals, including the Received Signal Strength (RSS), Time of Arrival (ToA) or Time Difference of Arrival (TDoA), Angle of Arrival (AoA)~\cite{overview2}; and
(2) data-driven approaches, which rely on collecting signal characteristics at known reference locations to construct a radio map and matching the observed signal patterns to the radio map~\cite{overview1}.
The first approach is straightforward but less accurate indoors than in outdoor settings due to the complex indoor signal propagation environments. In contrast, the data-driven approach leverages a detailed database of signal measurements collected at known reference locations (fingerprints) to estimate the target's location with greater accuracy.
However, collecting these signal measurements across the entire operational area requires expertise and is both time-consuming and resource-intensive, especially for large-scale deployments.

Crowdsourced data is increasingly regarded as a more viable and flexible alternative to traditional fingerprinting datasets, overcoming many of their limitations.
This approach relies on signal measurements collected from a large group of users across diverse devices and platforms. These measurements often include RSS, motion data, and various sensor readings.
Unlike traditional fingerprint datasets, which are collected by experts with precise location annotations, crowdsourced datasets typically lack location annotations. 
As a result, these {\em unlabeled data} cannot be directly utilized in existing fingerprinting-based localization frameworks and present challenges when integrated into localization.

In this paper, we propose a novel approach that leverages statistical information extracted from crowdsourced RSS data to estimate distances to access points (APs). These distance estimates are highly accurate and can be utilized for localization.
For illustration purposes, we introduce a localization framework that applies an improved trilateration algorithm to determine the target's location. While trilateration serves as a straightforward demonstration, our approach can be seamlessly integrated with other localization techniques to enhance accuracy and adaptability in real-world scenarios.
Experimental results show that our method achieves more accurate location estimations compared to the traditional Log-Distance Path Loss (LDPL) model.
The localization accuracy of our approach, which operates on crowdsourced datasets without location labels, is comparable to that of the $k$-Nearest Neighbor ($k$NN) estimation using the same dataset with ground truth location labels.


\subsection{Related Works}
\subsubsection{Fingerprinting-based Methods}
Fingerprinting-based methods typically involve two phases: (1) the offline phase, where a radio map is constructed with collected signal measurements (e.g., RSS or CSI) at known reference locations, and (2) the online phase, where real-time measurements are matched against the radio map to estimate the location~\cite{7905633}.
During the offline phase, received signals from nearby APs are measured at known reference locations across the area of interest (AOI). 
The measurements on received signals include RSS, AoA, and CSI~\cite{csi1, csi2}, which capture the features of the received signal at a given location. 
These measurements are typically represented and stored as a vector, with each element corresponding to a measurement relative to a specific AP.
The fingerprint dataset consists of the collection of these vectors, paired with the ground truth locations where they are obtained. 
The collected fingerprints are then used to construct a radio map, which contains the characteristics of measurements at various locations within the AOI.
The radio map contains each location's unique signal profile, providing a reference to estimate positions based on similarities.
During the online phase, the real-time signal measurements obtained by the user device will be sent to a positioning system. 
The positioning system estimates the position by comparing the obtained signal measurements with those previously surveyed on the radio map.
The positions are estimated using matching techniques, such as the nearest neighbor, which determines the location by identifying the closest matching signal measurements. 
Variants like $k$NN and weighted $k$NN improve the accuracy by considering multiple nearby reference locations and assigning weights based on similarity 
 ~\cite{832252, brunato2005statistical}.
Additionally, probabilistic methods, such as Maximum Likelihood Estimation (MLE)~\cite{youssef2005horus}, Bayesian methods~\cite{roos2002probabilistic, ladd2002robotics}, and Hidden Markov Models (HMM)~\cite{8747490}, have been used to estimate the location.
To enhance localization performance, machine learning approaches, including kernel estimation, clustering~\cite{5425278}, and support vector machines (SVM)~\cite{4079213}, are employed to infer the target location.
More recently, several advanced machine learning algorithms have successfully improved indoor localization accuracy and efficiency.
Recent studies have explored and proposed several neural network structures, including Convolutional Neural Networks (CNN)~\cite{yan2021device, 9773170, rm2} and Recurrent Neural Networks (RNN)~\cite{9109805}, to construct radio maps and estimate position.
Although several fingerprinting-based approaches achieve high accuracy, they all suffer from the following fundamental limitations:
\begin{itemize}
    \item Constructing accurate radio maps requires extensive signal measurements with precise location labels, making site surveys time-consuming, costly, and expertise-dependent. Thus, deploying fingerprinting-based methods in large or complex environments is particularly challenging, especially in commercial spaces like malls, airports, and factories.
    
    \item The radio map is highly sensitive to environmental changes, as signal characteristics like RSS and CSI vary with factors like object movement, furniture rearrangement, and human activity. Since it is built on measurements taken at the site survey, any changes in the environment can degrade its accuracy, leading to mismatches with real-time data.
    
\end{itemize}

\subsubsection{Calibration-Free Methods}
Calibration-free methods in wireless fingerprinting localization aim to bypass the traditional, labor-intensive site survey phase, where the environment is systematically scanned to collect reference signal fingerprints at known locations. 
Instead, these methods either reduce or completely eliminate the need for manual calibration by leveraging alternative data sources, automation, or adaptive algorithms.
Broadly, calibration-free methods can be categorized into two types: (1) those that utilize additional sensors, such as accelerometers, cameras, or compasses, to aid localization, and (2) those that operate purely based on wireless signals without requiring extra hardware.
In this work, we focus on the latter category, which relies on simulated data or machine learning techniques to construct a radio map and dynamically adapt to environmental changes. 
Floor plans and signal propagation models enable the generation of radio maps without requiring extensive on-site measurements.
Specifically, the floor plan is used to construct a 3D representation of the AOI, where ray-tracing simulations are performed to model signal behavior~\cite{8445928, 5297405}.
The resulting dataset can supplement or even replace the labor-intensive site survey.
Some studies assume that path loss follows statistical propagation models. The model parameters, such as the path loss exponent, are first estimated using a limited set of RSS measurements collected at known locations~\cite{zeytinci2013location, 4151127, Hu_2015}. During the online phase, these propagation models with learned parameters are used to estimate the distances to APs.
The estimated distances to at least three APs are then utilized to determine the target's position using trilateration.
However, empirical propagation models may deviate from real-world conditions due to environmental variations and unpredictable signal behaviors.
In addition, interpolation and extrapolation techniques are applied to increase the diversity of fingerprints, which enables the creation of a more comprehensive radio map with fewer actual measurements~\cite{talvitie_distance-based_2015}.
This approach aims to reduce the need for extensive site surveys and generates additional fingerprints between measured locations.
Recently, Generative Adversarial Networks (GANs) in machine learning have been proposed to synthesize RSS data in areas lacking coverage\cite{8891678}. 
By learning the underlying distribution of collected signal data, GANs can generate realistic radio signal measurements at unmeasured locations.
This approach reduces the need for extensive manual data collection and overall improves radio map coverage.
Furthermore, some work treats the radio map from a computer vision perspective, where super-resolution techniques are applied to enhance the resolution of the created radio map~\cite{9773170}.
Another line of research incorporates additional sensors to provide supplementary information that enhances the localization process. This additional sensor data improves the accuracy and robustness of fingerprinting-based localization systems.
For instance, motion sensors and accelerometers can estimate the distance traveled between two measurements, allowing the system to track the device's trajectory and refine location estimates. 
This additional distance information, combined with the fingerprint dataset, can be used in conjunction with multidimensional scaling to estimate location more accurately.
Indoor {\em landmarks} with known locations, including physical objects or features such as elevators, stairs, and doors~\cite{chen2015fusion, 7265092}, or unique signal sources (e.g., sniffers)~\cite{1356987}, serve as fixed reference points within the environment to enhance localization accuracy.
By cross-referencing the target's location with the known locations of these landmarks, both localization accuracy and robustness are improved.
Another approach utilizes measurement and location information over time~\cite{7949009, 7397867, 6568907}, assuming that the target moves from a known initial location and aims to track its trajectory. 
The signal measurements and sensor readings along the movement are combined to estimate the target's trajectory.
Fingerprinting-based methods provide location estimations at specific locations, while motion estimation techniques, such as accelerometers, track movement and orientation between these locations. 
Typically, techniques such as Kalman filters, extended Kalman filters, and particle filters are used to estimate the position over time.
By integrating the location information with continuous trajectory information, these methods achieve smoother, more accurate location tracking. 


\subsubsection{Geometric Approaches}
Another relevant line of work in indoor localization is the one that does not require site surveys and {\em labeled fingerprints} in the operational area.
These methods are primarily range-based, leveraging signal attenuation or time delay to estimate distances to APs using radio propagation models or ToA measurements. 
Once the distances to multiple APs are estimated, the target's position can be determined through triangulation, which calculates the intersection of distance-based regions to approximate the target's location.
A key advantage of geometric approaches is their ability to leverage crowdsourced data, which consists of location-related information collected from a large number of user-operated devices.
The data collection process involves the participation of users who contribute various types of data, such as RSS~\cite{calibration-free_2020}. 
Since the exact locations where crowdsourced data are obtained remains unknown, the crowdsourced data can be considered as {\em unlabeled fingerprints}.
Some studies propose LDPL-based conversion to transform unlabeled RSS measurements into distance estimates to WiFi APs~\cite{radar, calibration-free_2020}. To enhance accuracy, LDPL-based conversion has been combined with an AP alignment approach, as introduced in~\cite{koo2012unsupervised, plmodel-refine}, which adjusts for variations in AP coverage and improves the accuracy of distance estimates.
However, the localization performance of geometric approaches is often affected by environmental factors such as multipath effects and signal attenuation, particularly in indoor environments. These factors introduce distortions in RSS measurements and distance estimates, leading to increased localization errors.


\subsection{Contributions}
This paper introduces a novel approach to estimating distances to APs using statistical information derived from crowdsourced RSS data. 
Building on these accurate distance estimations, we propose a localization framework that operates effectively without requiring ground truth location labels. 
Remarkably, our approach's localization accuracy, which utilizes unlabeled data, is comparable to that of the $k$NN algorithm~\cite{7883033}, which relies on fingerprints with location labels.
The main contributions of this paper are outlined as follows:
\begin{itemize}
    \item 
    We analyze the statistical information of the crowdsourced data and propose a CDF-based algorithm for estimating the distances to APs. 
    Unlike existing methods, our approach does not rely on a specific path loss model and achieves significantly higher accuracy in complex indoor environments with multipath propagation and shadowing.
    These precise distance estimations can further enhance the performance of other localization frameworks that depend on range-based techniques.
    

    \item 
    We propose a novel localization framework based on the estimated distances to determine the target's location. Experimental results demonstrate that the localization accuracy achieved by our framework without ground truth location labels is comparable to that of the $k$NN algorithm, which relies on fingerprints with ground truth location labels. With comparable performance, our approach eliminates the need for labor-intensive site surveys, making localization more efficient and scalable.

    \item We generate crowdsourced data through extensive ray-tracing simulations using Remcom's Wireless InSite, a commercial ray-tracing software~\cite{Remcom2024}. We have also made our datasets publicly available~\cite{croudrt}. To evaluate the robustness of our framework, we evaluate its performance across different operating frequencies and receiver heights. Our method demonstrates strong generalization performance, successfully localizing even when crowdsourced data is collected at one frequency or height and tested at another. These results highlight the adaptability of our approach to diverse real-world scenarios with varying operating conditions.
    
    
\end{itemize}
This paper is organized as follows. Section II describes the system model.
Section III presents the distance estimation techniques used in the proposed method.
Section IV details the localization framework and methodology. 
Section V discusses the experimental results and performance evaluations. 
Section VI concludes the paper and explores potential future research directions.

\section{System Setup}

\begin{figure}
    \centering
    \includegraphics[width=0.45\textwidth]{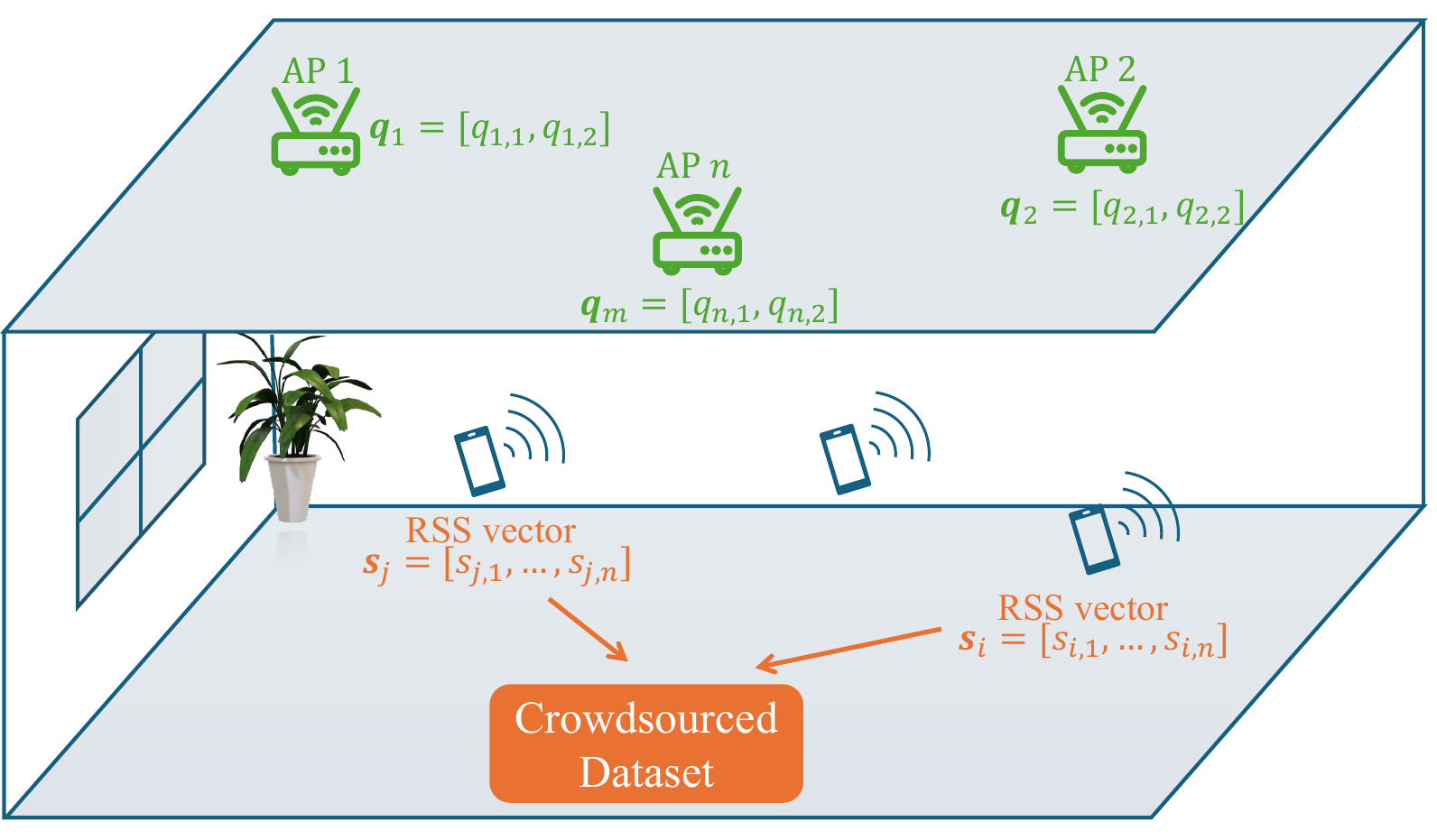}
    \caption{The simplified visualization of the system model.}
    \label{sys}
\end{figure}

We consider the indoor localization problem in two-dimensional space within the AOI $\ncalA \in \nbbR^2$, where $n$ APs periodically transmit beacon signals.
\figref{sys} illustrates a simplified, single-room scenario for clarity; however, our system setup is more general and considers more realistic, complex indoor environments involving multiple rooms.
The locations of the APs are known and represented by the $n \times 2$ matrix $Q = \left[\nbq_1, \nbq_2, \dots \nbq_n\right]$, where $\nbq_j = \left[q_{j,1}, q_{j,2}\right]^\intercal$ denotes the two-dimensional coordinate of the $j$-th AP. 
The crowdsourced RSS dataset contains $m$ RSS vectors collected by mobile devices at locations within the AOI, represented as $\ncalS = \left\{\nbs_1, \nbs_2, \dots, \nbs_m\right\}$. 
Each RSS vector contains RSS measurements from all $n$ APs, denoted by $\nbs_i = \left[s_{i,1}, s_{i,2}, \dots, s_{i,n}\right]$.
However, the exact locations where these $m$ vectors are measured remain unknown. We represent the ground truth locations as $\nbp_1, \nbp_2, \dots, \nbp_m$, where $\nbp_m = \left[p_{m,1}, p_{m,2}\right]^\intercal$ is the projected two-dimensional coordinate.
With the known locations of the APs and the crowdsourced data $\ncalS$, we aim to develop a localization algorithm to estimate the target's location $\nbp$ using the real-time RSS measurements $\nbs$ obtained at that location. 
Unlike some studies on positioning in wireless sensor networks, which assume a portion of crowdsourced data is collected at known locations, our work does not rely on location annotations.
We model $\nbp_1, \nbp_2, \dots, \nbp_m$ as realizations of a random variable $\nbP$ with a known PDF $f_\nbP(\nbp)$.
This means we assume that the underlying statistical distribution of the locations where crowdsourced RSS measurements are obtained is known.
We estimate the distances to the APs by leveraging the statistical information from both these locations and the crowdsourced RSS measurements. 
Specifically, with the collected crowdsourced data $\ncalS$ and the CDF $F_\nbP(\nbp)$, we map the RSS values to the distances to APs. 
The localization error is characterized by $e = |\nbp - \tilde{\nbp}|$, where $\nbp$ is the ground truth location of the target and $\tilde{\nbp}$ denotes the estimated location.

\section{RSS-Based Distance Estimation} \label{sec:pre}
We start with one-dimensional examples to build an intuitive understanding of our method of distance estimates in a controlled setting. 
The 1D case clearly illustrates the statistical relationship between RSS values and distances while reducing the complexity introduced by additional spatial dimensions.
Assuming the AP is located at the origin without loss of generality, we measure RSS values at various locations along the $x$-axis within the distance $[d_0, d_{\max}]$ to the origin.
These RSS measurements are denoted as $s_1, s_2, \dots, s_n$. 
However, the exact locations where these measurements are obtained remain unknown, and we represent their true distances to the AP as $d_1, d_2, \dots, d_n$.
The goal is twofold: (1) reconstruct the relationship between RSSs and distances, and (2) establish a mapping to estimate the distance to the AP using the RSS measurement. 
This mapping from RSS measurements to the distances is derived using statistical information about the measurement locations and the observed RSS values.

Our motivation is to first illustrate the fundamental principles of our approach in a controlled setting before extending it to more complex and realistic environments. 
We use a stochastic channel model that accounts for large-scale propagation effects, including path loss and shadowing.
The received signal power $P_r$ at distance $d$ to the transmitter is modeled as~\cite{goldsmith2005wireless, mimobook}:
\begin{align}
    P_r = P_t + K - 10 \gamma \log_{10}\frac{d}{d_0} - \chi \, {\rm (dB)},
\end{align}
where $P_t$ is the transmitted signal power, $K$ represents the path gain at the reference distance $d = d_0$, measured as $P_r / P_t$ in decibel scale, $\gamma$ is the path loss exponent, characterizing the rate of signal attenuation with distance, $d$ is the distance between the transmitter and the receiver, and $\chi$ is a random variable representing shadowing, typically modeled as a Gaussian random variable with zero mean and variance $\sigma_{\chi}^2$ with the probability density function (PDF)
\begin{align}
    f_{\chi}(\xi) = \frac{1}{\sqrt{2 \pi} \sigma_\chi} \exp \! \left( - \frac{\xi^2}{2 \sigma_\chi^2 }\right).
\end{align}
The shadowing effect captures the random variations in received signal power at a given distance caused by blockages and obstacles in the signal path. Extensive studies in the literature have examined the empirical autocorrelation function of the process of shadow fading process over distance.
The most widely used analytical model for this function, originally proposed by Gudmundson based on empirical measurements, assumes that shadowing follows a first-order Gudmundson process~\cite{gudmundson1991correlation}.
In this model, the covariance of shadow fading between two points separated by a distance $\delta$ is given by:
\begin{align}
    A(\delta) = \nbbE \left\{ \left(\chi(d) - \mu_\chi\right) \left(\chi(d + \delta) - \mu_\chi\right) \right\} = \sigma_\chi^2 \rho^{\delta / D},
\end{align}
where $\rho$ is the normalized covariance between two points separated by a fixed distance $D$, and $\sigma_\chi^2$ represents the variance of the shadow fading. 
To simplify the model and eliminate its empirical dependence~\cite{goldsmith2005wireless}, $\rho_D$ is set to $1/e$ for a distance $D = X_c$, which which leads to the following expression:
\begin{align}
    A(\delta) = \sigma_\chi^2 e^{-\delta / X_c},
\end{align}
where $X_c$ is the decorrelation distance.
It is important to emphasize that the specific forms of the correlation function, the path-loss function, and the assumptions about shadowing presented here are chosen for illustrative purposes. Our proposed method is not dependent on these particular assumptions and does not require explicit knowledge of them. Consequently, our approach remains fully applicable under different modeling conditions and can be easily adapted to other scenarios as needed.

We simulate the indoor environment using the parameters listed in Table~\ref{tab:sim_params}. The received signal power, considering the combined effects of path loss and shadowing, is plotted in decibel scale as a function of the transmitter-receiver distance in~\figref{c1}.
\begin{table}[!ht]
\centering
\caption{Simulation Parameters for Indoor Environment}
\begin{tabular}{|c|c|}
\hline
\textbf{Parameter}        & \textbf{Value}          \\ \hline
Decorrelation distance $X_c$ & $10 \, \nrmm$      \\ \hline
Shadow fading variance $\sigma_{\chi}$ & $10 \, {\rm dB}$    \\ \hline
Minimum distance $d_0$   & $2 \, \nrmm$       \\ \hline
Maximum distance $d_{\max}$ & $25 \, \nrmm$     \\ \hline
Path loss exponent $\gamma$ & $3$                   \\ \hline
Transmitted signal power $P_t$ & $0 \,{\rm dBm}$        \\ \hline
\end{tabular}
\label{tab:sim_params}
\end{table}

\begin{remark}
Readers might expect scatter plots of RSS measurements as a function of distance with significant variance, where RSS measurements are scattered around the average RSS. 
This high variance occurs because RSS measurements are obtained in a two-dimensional space along different directions relative to the AP. 
For instance, measurements taken at the same distance from the AP correspond to points lying on a circle centered at the AP, which are not necessarily close to each other (unlike what the distance from the AP might suggest).
It is well known that the correlation exhibited by shadowing depends on separation distance, i.e., closer distances tend to stronger spatial correlation in shadowing~\cite{goldsmith2005wireless,gudmundson1991correlation}.
When RSS values from all directions are combined in a scatter plot as a function of distance from the AP, the high variance hides the spatial correlation, especially when measurements come from multiple APs.
This occurs because the separation distance of RSS measurements cannot be fully inferred from the distance to the AP.
When RSS measurements are plotted along a specific direction, as shown in~\figref{c1}, the spatial correlation becomes more evident.
If we combine RSS measurements from multiple directions, the resulting scatter plot resembles multiple realizations of the RSS measurements from individual directions. This scatter plot will have high variance, which aligns with the expectation.
\end{remark}


To understand the method of transforming RSS measurements into distances to APs, we will examine three different scenarios with different levels of statistical information available about the locations of crowdsourced measurements.
We begin with a simple but illustrative setting where measurements are obtained at locations evenly spaced at fixed intervals (Case 1). In this scenario, distances can be estimated using a straightforward approach that relies on ordering the RSS measurements.
Next, we relax the assumption of fixed intervals and consider RSS measurements obtained at locations that are uniformly at random sampled within the same interval (Case 2). The distances to AP can still be effectively estimated using the same ordering approach as in Case 1.
Finally, we further relax the assumption of a uniform distribution and extend the distribution of the locations to an arbitrary distribution (Case 3). 
The uneven location sampling limits the effectiveness of the ordering approach.
To address this, we utilize the statistical information of the locations and RSS measurements to propose a CDF-based transforming method.

\begin{remark}
We assume our prior knowledge is the distribution of locations where crowdsourced data is obtained. 
This knowledge can be derived from historical data, such as the trajectories of mobile phone users whose positions are tracked.
Some studies on human mobility patterns have demonstrated that human trajectories have a high degree of spatial regularity~\cite{gonzalez2008understanding}. For example, historical data may reveal that individuals spend five times more time in their office cubicles than in common areas such as hallways or lounges. Moreover, physical constraints also contribute to prior information. Map-based priors capture the fact that targets cannot occupy physically impossible locations, such as walls, restricted zones, or areas outside the building. 
In some environments, functional zones (e.g., seating areas, entrances, or equipment rooms) may also have different likelihoods of occupancy, which further informs the prior distribution.
\end{remark}


\subsection{Case 1: Uniformly Spaced}

\begin{figure}
    \centering
    \includegraphics[width=0.45\textwidth]{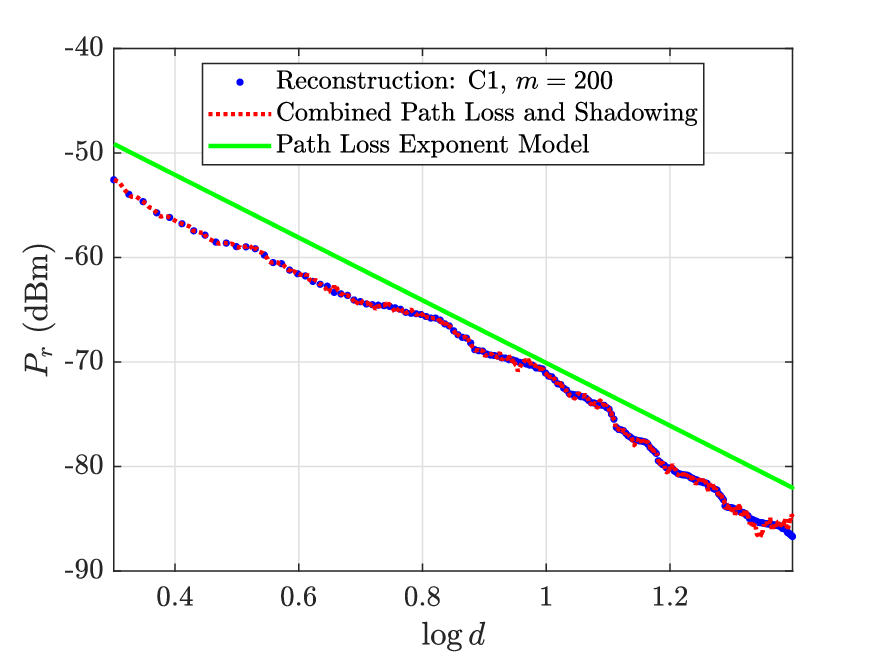}
    \caption{Illustration of Case 1: the received signal power as a function of distance. The blue dots represent the reconstructed RSS to distance relationship using $m = 200$ samples. The red line indicates the simulated RSS along the positive $x$-axis using a stochastic channel model that incorporates both path loss and shadowing effects. For comparison, the green solid line shows the theoretical path loss with exponent $\gamma = 3$.}
    \label{c1}
\end{figure}

In the first case, we consider RSS measurements obtained at locations that are uniformly spaced within the interval $[d_0, d_{\max}]$.
These locations lie along the positive $x$ axis and at distances of $d_1, d_2, \dots, d_m$ to the AP placed at the origin, where each distance is given by $d_i = d_0 + \frac{i}{m + 1} d_{\max} $.
However, the exact locations where the RSS measurements are obtained remain unknown.
We aim to estimate these unknown locations corresponding to the RSS measurements $s_1, s_2, \dots, s_m$.
Due to shadow fading, we know that RSS does not strictly decrease with distance; however, it generally trends downward as the receiver moves farther from the transmitter due to path loss. 
This relationship provides valuable insight, as RSS can be approximated as a non-increasing function of distance, which serves as useful information in distance estimation and localization.
We use this information to estimate the distances to the AP by arranging the RSS measurements in descending order, $s_{(1)} \geq s_{(2)} \geq \dots, \geq s_{(m)}$, and associating each RSS measurements $s_{(i)}$ with its corresponding distance estimation $\hat{d}_i = d_0 + \frac{i}{m + 1} d_{\max} $.
This ordering approach establishes a mapping from RSS measurements to distances and estimates the unknown locations where the measurements were obtained.
We applied this method to the simulated RSS measurements using the parameters listed in Table~\ref{tab:sim_params}.
The results, illustrated in~\figref{c1}, show that the reconstructed RSS-to-distance relationship obtained with the ordering method (blue dots) closely aligns with the simulated RSS (red dashed line), demonstrating that the ordering method effectively captures the variations in received signal power over distance. 
However, since RSS does not strictly decrease with distance due to shadow fading, the reconstructed RSS-to-distance relationship deviates from the ground truth when $\log d$ is above $1$. 

The fundamental limitation of estimating distance using RSS measurements arises when different locations yield the same RSS values, making it difficult to distinguish between them.
Mathematically, distance estimation algorithms using the RSS can be conceptualized as attempts to find the inverse function of the measurement, where the RSS is mapped back to its corresponding location. However, without additional measurements beyond RSS, even if the RSS-to-distance relationship is known, accurately determining the exact location corresponding to a specific RSS value remains impossible.
To formalize this limitation, we present the following lemma from real analysis.


\begin{remark}
A continuous function does not necessarily have an inverse function. If the inverse function exists, the continuous function must be strictly monotonic.
\end{remark}
Due to the inherent limits in the mapping function, this inverse mapping may not always yield the correct solution.
When identical RSS values can be observed at multiple locations, the optimal inverse mapping without any localization error may not exist. 
Our approach aims to approximate the inverse function.
Although it cannot fully reconstruct the many-to-one aspects of the RSS-to-distance relationship, it remains highly effective for localization purposes.

\subsection{Case 2: Uniformly Distributed}
\begin{figure}
    \centering
    \includegraphics[width=0.45\textwidth]{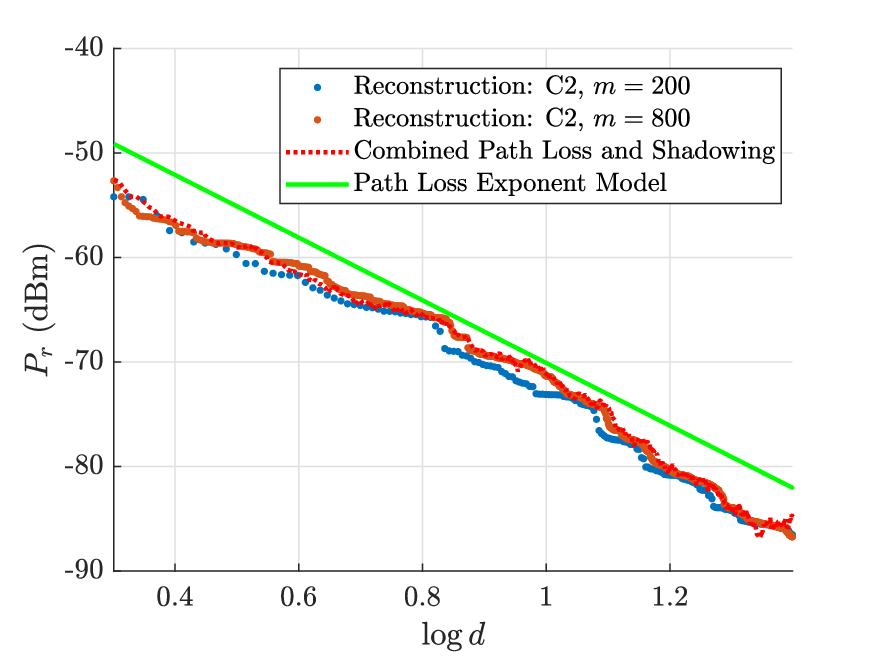}
    \caption{Illustration of Case 2: the received signal power as a function of distance. The RSS measurements are obtained at locations randomly sampled uniformly over the positive $x$-axis in the interval $[d_0, d_{\max}]$. The distances are estimated using crowdsourced RSS measurement datasets with two different sizes: $m=200$ (blue dots) and $m=800$ (orange dots). The red line represents the true received signal power, and the green solid line corresponds to the theoretical path loss with exponent $\gamma = 3$ for reference.}
    \label{c2}
\end{figure}

We now relax the assumption of uniformly spaced locations and instead, consider RSS measurements obtained at locations that are uniformly at random selected along the positive $x$-axis.
These locations lie within the distance interval $[d_0, d_{\max}]$ to the AP at the origin and are represented as $d_1, d_2, \dots, d_m$. Each instance of the distance is independently sampled from the uniform distribution with support $[d_0, d_{\max}]$.
As before, the exact locations where the RSS measurements $s_1, s_2, \dots, s_m$ are obtained remain unknown. 
Our goal is to estimate the unknown locations where the RSS measurements are obtained. 
We apply the ordering method by arranging the RSS measurements in descending order $s_{(1)} \geq s_{(2)} \geq \dots, \geq s_{(m)}$.
For the $i$-th ordered RSS measurement $s_{(i)}$, the distance estimation is calculated as $\tilde{d}_i = d_0 + \frac{i}{m+1} d_{\max}$. 
In Appendix~\ref{app-c2}, we prove that when the RSS is a monotonic function of distance, the proposed distance estimation method is unbiased. The variance of the estimation error under this scenario is also derived. When RSS is not a monotonic function, the derived variance serves as a lower bound. Please see Appendix~\ref{app-c2} for more details.
We demonstrate the reconstructed RSS-to-distance relationship using the ordering method in~\figref{c2}. 
The result shows that the reconstructed relationship closely follows the ground truth RSS as a function of distance (red dashed line).
Furthermore, increasing the sample size $m$ from $200$ to $800$ improves the accuracy of the location estimation, as evidenced by the closer alignment of the orange dots with the red dashed line. 


\subsection{Case 3: Randomly Distributed}
\begin{figure}
    \centering
    \includegraphics[width=0.45\textwidth]{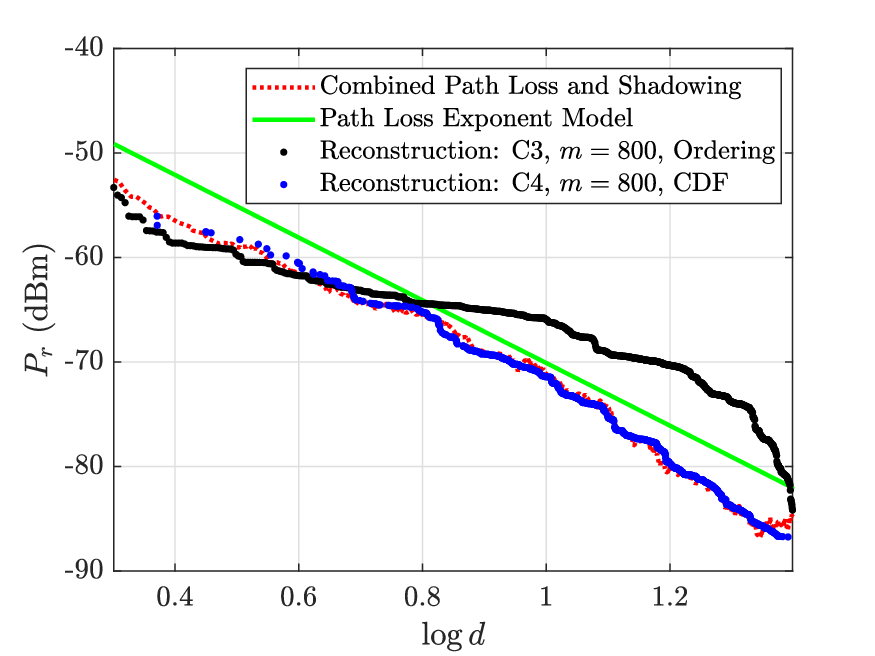}
    \caption{Illustration of Case 3: the received signal power as a function of distance with $m = 800$ samples.
    The RSS-to-distance relationship is reconstructed using the ordering method (black dots) and the CDF method (blue dots), respectively.
    The RSS measurements are obtained at locations along the positive $x$-axis, where the underlying distance distribution follows a beta distribution with parameters $\alpha = 2$ and $\beta = 2$, scaled to the interval $[d_0, d_{\max}]$. 
    The red line represents the true path loss, and the green solid line corresponds to the theoretical path loss with exponent $\gamma = 3$ for reference.}
    \label{c3}
\end{figure}

We further relax the assumption that the measurements are collected at locations sampled from a uniform distribution. 
Instead, we consider crowdsourced RSS measurements obtained at locations along the positive $x$-axis, where the locations are sampled from an arbitrary distribution with finite support $[d_0, d_{\max}]$.
These sampled locations are at distances $d_1, d_2, \dots, d_m$ from the origin, where each $d_i$ an independent realization of the random variable $D$, drawn from a distribution with a known PDF $f_D(d)$.
The RSS measurements obtained at these locations are represented as $s_1, s_2, \dots, s_m$. 
The goal is to estimate the locations associated with RSS measurements without knowing the distances $d_1, d_2, \dots, d_m$. 
We first apply the previously used ordering method and present the results as black points plotted in~\figref{c3}.
However, since the locations are not sampled from a uniform distribution, simply arranging the RSS measurements in decreasing order and associating them with evenly spaced distances $\tilde{d}_i = d_0 + \frac{i}{m+1} d_{\max}$ does not work.
The uniform spacing of estimated distances does not accurately represent the true distribution of the measurement locations.
To overcome this limitation, we utilize the statistical information of both the locations and the RSS measurements and introduce a CDF-based transformation method.
In this approach, we treat the distance to the AP as a random variable $D$ and the RSS as another random variable $S = g(D)$, where $S$ is a function of $D$.
The basic properties related to random variables and their transformations are provided next to help in understanding the analyses conducted in this paper.

\subsection{The PDF Transformation}
Let $D$ be a random variable with a known PDF $f_D(d)$, and $S = g(D)$ represents the second random variable as a function of $D$.
The distribution of $S$ can be derived from the known distribution of $D$ using the method of transformations, as outlined in the following lemma:
\begin{lemma} \label{lem1}
Suppose that $D$ is a continuous random variable and $g:\nbbR \rightarrow \nbbR$ is a strictly monotonic differentiable function. Let $S = g(D)$, and the PDF of $S$ is given by
\begin{align}
    f_S\!\left(s\right) = f_D\!\left(g^{-1}\! \left( s \right)\right) \left| \frac{\nrmd}{\nrmd s} g^{-1}\! \left(s\right) \right|,
\end{align}
where $g^{-1}(\cdot)$ is the inverse function of $g(\cdot)$.
\end{lemma}

\begin{IEEEproof}
Please see~\cite{pishro2014introduction} for a detailed proof.
\end{IEEEproof}

Lemma~\ref{lem1} establishes the relationship between the PDF of the random variable $D$, the PDF of its transformation $S = g(D)$, and the transformation function $g$. 
This relationship can also be expressed in terms of CDF as
\begin{align} \label{cdf-trans}
    F_S\!\left(s\right) = 1 - F_D\!\left(g^{-1}\!\left(s\right)\right),
\end{align}
where $g:\nbbR \rightarrow \nbbR$ is a strictly monotonic decreasing function of $D$, $F_D$ is the CDF of random variable $D$, and $F_S$ is the CDF of random variable $S$. 

In statistical learning, we may encounter scenarios where we have unlabeled values $s_1, s_2,\dots, s_m$.
However, the transformation function $f$ and the actual values of $D$, denoted as $d_1, d_2,\dots, d_m$, remain unknown. 
In this case, it is challenging to estimate the corresponding values $d_1, d_2,\dots, d_m$ for the observed data $s_1, s_2,\dots, s_m$.
This is the common situation in various applications of unsupervised learning, such as the distance estimation problem discussed in this paper, where we have a collection of crowdsourced RSS measurements (considered as $s_1, s_2, \dots, s_m$) and aim to estimate the distances to APs (denoted as $d_1, d_2,\dots, d_m$).
Before introducing our proposed distance estimation method, we first review existing approaches in the literature and their underlying assumptions.

\subsection{RSS-to-Distance Estimation}
In the literature on localization using crowdsourced RSS data, several studies have focused on estimating the distances to APs based on the LDPL model~\cite{koo2012unsupervised, calibration-free_2020}.
They establish a linear mapping from RSS measurements to the logarithm of distances using the maximum and minimum RSS collected in the crowdsourced dataset.
Specifically, the minimum RSS value is mapped to a known maximum coverage distance $L_{\rm ref}$, while the maximum RSS value is mapped to a zero distance.
Intermediate RSS values are then linearly scaled to distances within the range $[0, L_{\rm ref}]$.
This approach relies on the underlying assumption that each AP operates at an identical transmission power level and shares a common maximum coverage radius.
However, when the coverage radii of APs differ due to variations in transmission power or environmental conditions, applying the same coverage distance $L_{\rm ref}$ for all APs becomes inaccurate and can lead to significant localization errors.
To address variability in AP coverage radii, the model proposed in~\cite{plmodel-refine} introduces an AP alignment method that calibrates the effective coverage radius.
Their analysis of the crowdsourced dataset revealed that the number of valid RSS measurements is approximately proportional to the square of each AP's radio coverage radius. 
This observation directly follows from the assumption that crowdsourced measurements are obtained at random locations uniformly distributed within the AOI, implying that the number of RSS measurements associated with each AP is proportional to the size of its coverage area.
These distance estimation methods have several potential limitations:
(1) The LDPL model assumes a simplified, single-slope relationship between RSS and distance for modeling path loss. However, this assumption may not accurately reflect real-world environments where obstacles, reflectors, and multipath propagation introduce significant complexities in signal behavior.
(2) The assumption that measurement locations are uniformly and randomly distributed within the AOI may not hold in practice. Non-uniform data collection can impact the accuracy of the calibration process and lead to biased distance estimations.
(3) This linear mapping between RSS and distance depends on the minimum and maximum RSS values collected in the database.
These values are highly susceptible to noise and can be distorted by environmental factors, device sensitivity, and signal interference. Such variability can lead to inaccurate distance estimations and undermine the reliability of the localization model.

Another line of work estimates the distances to APs using statistical propagation models. 
In these approaches, model parameters, such as the path loss exponent, are estimated using RSS data collected at known locations~\cite{zeytinci2013location, 4151127, Hu_2015}. Once the model parameters are learned, the statistical propagation model is applied to estimate the distances to APs based on the observed RSS measurements.
This approach to distance estimation relies heavily on accurately matching the statistical propagation model to the specific environment and the precision of the estimated model parameters. 
Moreover, the parameter estimation process requires RSS data to be collected at known locations, which fundamentally differs from the crowdsourced localization setting where measurement locations are typically unknown or unlabeled.

In this paper, we leverage the statistical distribution of crowdsourced RSS data to estimate distances. This approach eliminates the need for assumptions about specific propagation models and offers greater adaptability to the variability and complexity of indoor environments.
The collected RSS values are treated as samples from the random variable $S$, and we estimate the distance $\tilde{d}$ using the inverse function $g^{-1}(\cdot)$, given as
\begin{align}
    \tilde{d} = g^{-1}(s) = F_D^{-1} \! \left(1 - F_S \right( s\left)\right),
\end{align}
where $F_D^{-1}$ is the inverse of the CDF of $D$, $F_S(s)$ is the empirical CDF of $S$. 
This relationship directly follows from Equation~\eqref{cdf-trans}.
The CDF of $D$ is assumed to be known, and the empirical CDF of $S$ can be computed from the collected RSS data $s_1, s_2, \dots, s_m$.
We apply the CDF-based transformation method to estimate distances for crowdsourced RSS measurements collected at locations sampled from an arbitrary distribution with known PDF.
While the PDF is assumed to be known, its specific choice is not critical to our analysis, as the method remains applicable to arbitrary distributions.
The simulation results are presented in~\figref{c3}. We observe that our CDF-based method can accurately estimate distances even when the exact measurement locations are unknown and unevenly distributed.

\begin{remark}
When identical RSS values are observed at multiple locations, the estimated distance may deviate from the actual distance.
Even when the relationship between RSS and distance is known, accurately determining the exact location where a measurement is obtained among two or more possible locations is impossible without additional information.
This challenge is common to most localization algorithms, as identical RSS values can correspond to multiple locations. To mitigate this limitation, we employ clustering techniques in later sections and refine location estimation using more than three distance estimates.
\end{remark}

Our approach estimates the distances to APs by leveraging the statistical distribution of crowdsourced RSS measurements. By avoiding reliance on specific propagation models, this method is well-suited for indoor localization using crowdsourced RSS data, where measurements are collected without associated location labels.
Our distance estimation approach not only provides accurate distance predictions but can also be seamlessly integrated with other localization algorithms to enhance overall positioning accuracy. 

\section{CDF-Based Localization}
\begin{figure*}
	\centering
    \includegraphics[width=0.9\textwidth]{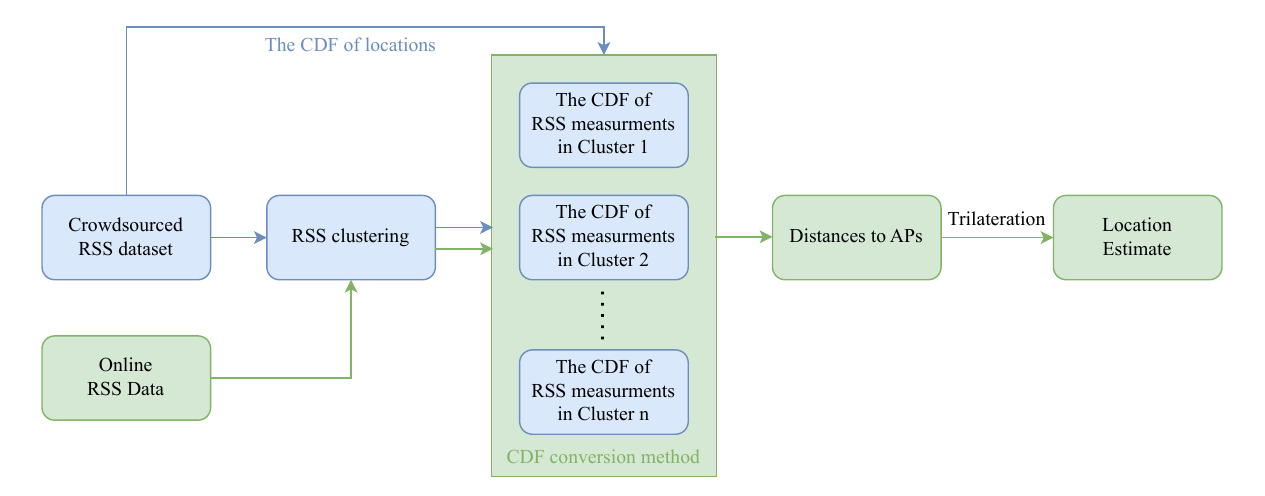}
	\caption{The system flowchart.}
    \label{flow}
\end{figure*}
In the previous section, we introduced an approach to estimating distances that utilizes the statistical properties of the locations and the RSS measurements.
Now, we apply our proposed distance estimation method to address the indoor localization problem using crowdsourced data, where the locations of the measurements are typically unknown. 
We propose a three-step framework for indoor localization using the crowdsourced RSS dataset:
\begin{itemize}
    \item {\bf RSS Vector Clustering}: We group the RSS measurements into clusters based on their similarity. This helps identify patterns and associate measurements that are likely obtained from nearby locations.
    \item {\bf Distance Estimation}: The CDF-based transformation method described in the previous section is applied to each cluster. This step converts the collected RSS data into estimated distances to the AP.
    \item {\bf Position Estimation}: The unknown positions are estimated using an enhanced trilateration algorithm. This improved algorithm utilizes the estimated distances to calculate the locations of the measurements.
\end{itemize}
The flowchart of the framework is illustrated in~\figref{flow}.
The localization performance of our framework is evaluated using RSS data generated by Ray-tracing software~\cite{Remcom2024}, incorporating a 3D model of the floor plan to simulate realistic signal propagation.
The results show that location estimates based on RSS data, even without ground truth location labels, achieve accuracy comparable to the $k$NN algorithm, which relies on known location labels. 
This result demonstrates that our approach effectively utilizes crowdsourced data for location estimation, achieving competitive accuracy even without ground truth location labels.
Notably, our framework can be seamlessly integrated with other localization techniques using the distance estimations between APs and targets.

\subsection{RSS Clustering}
Due to shared environmental conditions that affect signal propagation, wireless signals received at nearby locations are likely to be similar.
We aim to establish a relative location relationship between RSS measurements by first clustering the crowdsourced RSS vectors based on their similarity. 
We will explore more examples in the numerical results section, but for a fair comparison of our proposed location estimation method without location labels to the $k$NN algorithm with location labels, we temporarily exclude side information from our framework. 

We first divide the AOI into several spatial clusters, resulting in a partition of the AOI. For illustration, one example of this partitioning method is based on the locations of the APs, resulting in the Voronoi cells. Other partitioning methods can also be applied within our framework, such as K-means clustering or K-Voronoi cells (KVC). We will include these alternative methods in the numerical results section to explore their performance and compare their effectiveness in the context of our framework.
The Voronoi cells are defined as
\begin{align}
    V_i = \left\{ \nbx \in \ncalC : \left\| \nbx - \nbq_i\right\| \leq \left\| \nbx - \nbq_j\right\|, \forall k \in \left\{ n\right\}\right\},
\end{align}
where $V_i$ represents the Voronoi cell containing the $i$-th AP, $\nbq_i$ is the location of the $i$-th AP, $j \in \left\{ n\right\} = \left\{1, 2, \dots, n\right\}$ is the AP indices.
Since the locations within the Voronoi cell $V_i$ are closest to the $i$-th AP, the RSS from this AP is expected to be the strongest in this region.
Thus, we assign the RSS vectors with the largest value at the $i$-th entry to the Voronoi cell $V_i$.
We represent the set of RSS vectors labeled with $V_i$ as
\begin{align}
    \ncalS_i = \left\{\nbs_m : s_{m, i} > s_{m, j}, \forall j \in \left\{n \right\} \right\},
\end{align}
where $\nbs_m = \left[ s_{m, 1}, s_{m, 2}, \dots, s_{m, n}\right]$ is the RSS vector. The set $\ncalS_i$ includes all the RSS measurements with label $V_i$ that have the strongest received signal from the $j$-th AP.

We plot the ground truth locations and the clustering result based on the maximum RSS in~\figref{cell_a} and~\figref{cell_b}. The different colors represent the labels of the RSS vectors. Most of the RSS vectors are correctly labeled. 
The clustering results reveal that clustering based on the maximum RSS value provides an approximate indication of the area from which each RSS vector was obtained. 
The RSS measurements with the same label are obtained from the same area, where the received signals likely experience similar attenuation and propagation effects.

Using the clustering results, we apply our proposed CDF-based method to estimate the distances for the RSS vectors in each cluster. This approach overcomes the challenge of estimating distances across the AOI by breaking it into smaller subproblems for each Voronoi cell. Since the received signals within each Voronoi cell are likely to experience similar attenuation and propagation effects, they can share the same path loss model. This specific clustering method is used for illustration purposes and one can use other clustering methods, such as $K$-means and $K$ Voronoi cells. We present the localization performance when using the various clustering methods in the numerical result section.

\subsection{RSS Conversion}
Due to potential multipath effects and shadowing in the indoor environment, precisely estimating these distances based on the RSS is challenging.
We propose an RSS conversion method that leverages the statistical properties of crowdsourced data, allowing for more accurate distance estimations. 
First, we select a cluster of RSS vectors $\ncalS_i$.
These measurements are likely influenced by the same propagation environment, such as being within the same Voronoi cell $V_i$, where the received signal characteristics remain similar due to proximity to the same AP.
Without loss of generality, we assume that measurements are obtained from locations randomly distributed within the Voronoi cell according to some PDF $f_{\nbP}(\nbp)$.
The exact position where the RSS is obtained remains unknown.
Now, consider the $j$-th entry of the RSS vectors in $\ncalS_i$, representing the RSS from the $j$-th AP. 
We treat these RSS measurements as the random variable $S$, as described in Section~\ref{sec:pre}.
The empirical CDF of $S$ can be calculated based on the obtained RSS values.
The distances to the $j$-th AP can be thought of as 
random variable $D$ described in Section~\ref{sec:pre}. 
Since the locations are randomly distributed within $V_i$ with PDF $f_{\nbP}(\nbp)$, the CDF of the distance to the $j$-th AP can be calculated by sampling locations within $V_i$. The result is the CDF of random variable $D$.
As distance from the AP increases, the RSS value decreases. 
We can generally consider $S$ (the RSS value) as a monotonically decreasing function of $D$ (the distance). 
The decreasing function includes exponential path loss as described in the LDPL model but is not limited to it. 
Other functional forms may also be used to model the relationship between RSS and distance, accommodating a variety of propagation environments and attenuation behaviors. 
The conversion from RSS values to distances to APs is constructed using the CDFs of $D$ and $S$, given as
\begin{align}
    F_S\!\left(s\right) = 1 - F_D\!\left(g^{-1}\!\left(s\right)\right),
\end{align}
where $D$ represents the distance to APs, $S$ represents the RSS value.
We present the pseudo-algorithms of RSS conversion in Algorithm~\ref{alg:rss}.
\begin{algorithm}
\caption{RSS Conversion}\label{alg:rss}
\begin{algorithmic}
\Input $f_{\nbP}(\nbp), \left\{V_k\right\}_{k=1:n}, \left\{\ncalS_k\right\}_{k=1:n}, Q $
\Output $\left\{d_{i,j}\right\}_{i=1:m, j=1:n}$
\For{every Voronoi cell $k=1:n$}
\For{every AP $j=1:n$}
    \State Compute the CDF $F_{D_k}(d_k)$ based on $\nbq_j$, $V_k$ and $f_{\nbP}(\nbp)$
    \State Compute the CDF $F_{S_{kj}}(s_{kj})$ based on the $j$-th entry of RSS vectors in $\ncalS_k$
    \State $d_{i,j} = F^{-1}_{D_k}\!\left( 1 - F_{kj} \left( 
 s_{i,j}\right)\right)$ 
\EndFor
\EndFor
\end{algorithmic}
\end{algorithm}

In the real world, RSS does not strictly decrease with distance due to shadowing and multipath effects. To address this, our clustering method partitions the AOI and approximates the inverse function to estimate distances within each partition. 
The overall RSS-to-distance conversion across the AOI accommodates the non-monotonic behavior of RSS measurements.
This method achieves localization accuracy comparable to the $k$NN algorithm, which relies on fingerprints with ground-truth location labels. It also outperforms the LDPL-based approaches, as demonstrated in the numerical results section.

\subsection{Trilateration}
Using the estimated distances, we determine the location of the target. The unknown position is calculated using the trilateration method. We represent the locations of the APs as $Q = \left[ \nbq_1, \nbq_2, \dots, \nbq_n\right]$, where $\nbq_j$ denotes projected two-dimensional coordinate of the $j$-th AP.
The estimated distances between the target's location and each AP are denoted by $d_{1}, d_{2}, \dots, d_{n}$, respectively. 
This distance-location relationship is expressed by the following nonlinear equations:
\begin{align}
    \begin{cases}
    \begin{matrix}
        \left( p_{i,1} - q_{1,1} \right)^2 + \left( p_{i,2} - q_{1,2}\right)^2 = d_{1}^2, \\
        \left( p_{i,1} - q_{2,1} \right)^2 + \left( p_{i,2} - q_{2,2}\right)^2 = d_{2}^2, \\
        \vdots \\
        \left( p_{i,1} - q_{n,1} \right)^2 + \left( p_{i,2} - q_{n,2}\right)^2 = d_{n}^2.
    \end{matrix}
    \end{cases}
\end{align}
To simplify the calculation, we subtract the $n$-th equation from the first $n-1$ equations, resulting in a set of linear equations that relate the location of the target $\nbp = \left( p_{1}, p_{2}\right)$ to the known positions of APs and the distances between them. These linear equations can be written as
\begin{align}
\begin{aligned}
\begin{cases}
    \begin{aligned}
    2&\left(q_{1,1} - q_{n,1}\right)p_{1} + 2\left(q_{1,2} - q_{n,2}\right)p_{2} \\
    &= q_{1,1}^2 - q_{n,1}^2 +  q_{1,2}^2 - q_{n,2}^2 -  d_{1}^2 + d_{n}^2,
\end{aligned}\\
\begin{matrix}
    \begin{aligned}
    2&\left(q_{2,1} - q_{n,1}\right)p_{1} + 2\left(q_{2,2} - q_{n,2}\right)p_{2} \\
    &= q_{2,1}^2 - q_{n,1}^2 + q_{2,2}^2 - q_{n,2}^2 - d_{2}^2 + d_{n}^2,
\end{aligned}\\
\vdots
\end{matrix} \\
\begin{aligned}
    2&\left(q_{n-1,1} - q_{n,1}\right)p_{1} + 2\left(q_{n-1,2} - q_{n,2}\right)p_{2} \\
    &= q_{n-1,1}^2 - q_{n,1}^2 + q_{n-1,2}^2 - q_{n,2}^2 - d_{n-1}^2 + d_{n}^2.
\end{aligned}
\end{cases}
\end{aligned}
\end{align}
The above equation can be rewritten in matrix form as:
\begin{align}
    \nbA\nbx = \nbb,
\end{align}
where
\begin{gather}
    \nbx = \left[ p_{i,1}, p_{i,2}\right]^\intercal,
\end{gather}
\begin{gather}
\vspace{-1 in}
    \nbA = 
    \begin{bmatrix}
    2(q_{1,1} - q_{n,1}) & 2(q_{1,2} - q_{n,2}) \\
    \vdots & \vdots \\
    2(q_{n-1,1} - q_{n,1}) & 2(q_{n-1,2} - q_{n,2})
    \end{bmatrix}, \\
    \nbb \! = \!
    \begin{bmatrix}
         q_{1,1}^2 - q_{n,1}^2 + q_{1,2}^2 - q_{n,2}^2 -  d_{i,2}^2 + d_{i,n}^2 \\
        \vdots\\
        q_{n-1,1}^2 - q_{n,1}^2 + q_{n-1,2}^2 - q_{n,2}^2 - d_{i,n-1}^2 + d_{i,n}^2
    \end{bmatrix}.
\end{gather}
The unknown location of the target can then be estimated by solving the following equation:
\begin{align}
    \tilde{\nbp} = \left( \nbA^\intercal \nbA \right)^{-1} \nbA^\intercal \nbb.
\end{align}
This result extends the trilateration and provides a more robust estimate of the target's position by leveraging more than three distances to APs.

\section{Numerical Results}
\begin{figure*}
	\centering
	\subfigure[]{
    \begin{minipage}[t]{0.60\textwidth}
    \centering
    \includegraphics[width=\textwidth]{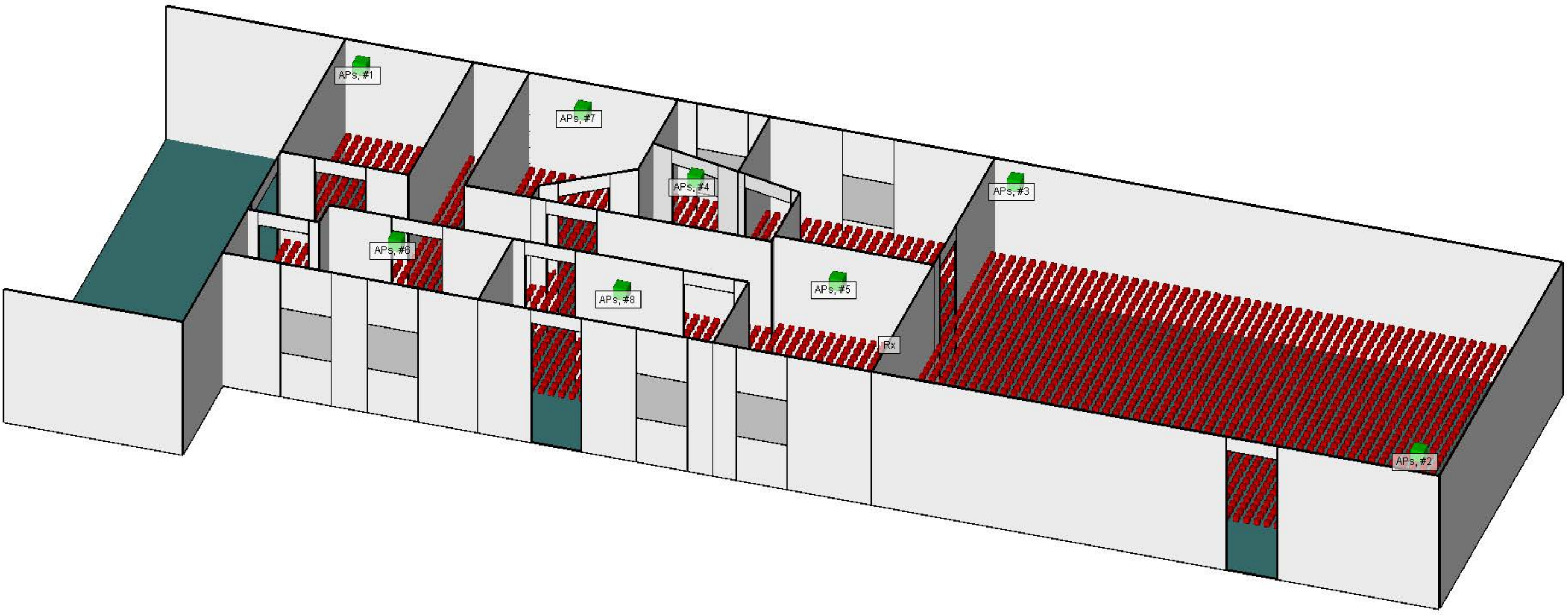}
    \label{fp1}
    \end{minipage}
    }
    \subfigure[]{
    \begin{minipage}[t]{0.35\textwidth}
    \centering
    \includegraphics[width=\textwidth]{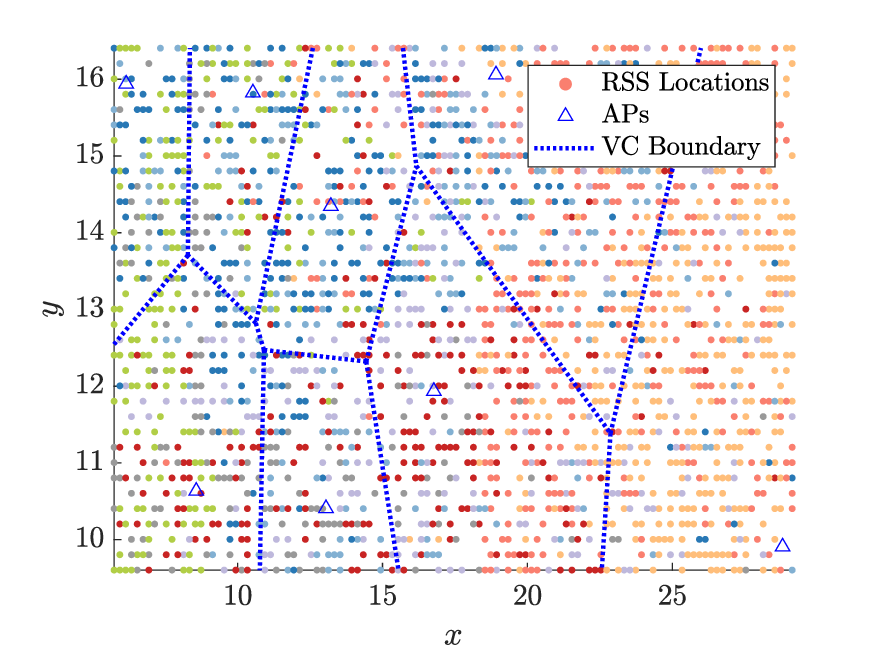}
    \label{cell_a}
    \end{minipage}
    }\\
    \subfigure[]{
    \begin{minipage}[t]{0.50\textwidth}
    \centering
    \includegraphics[width=\textwidth]{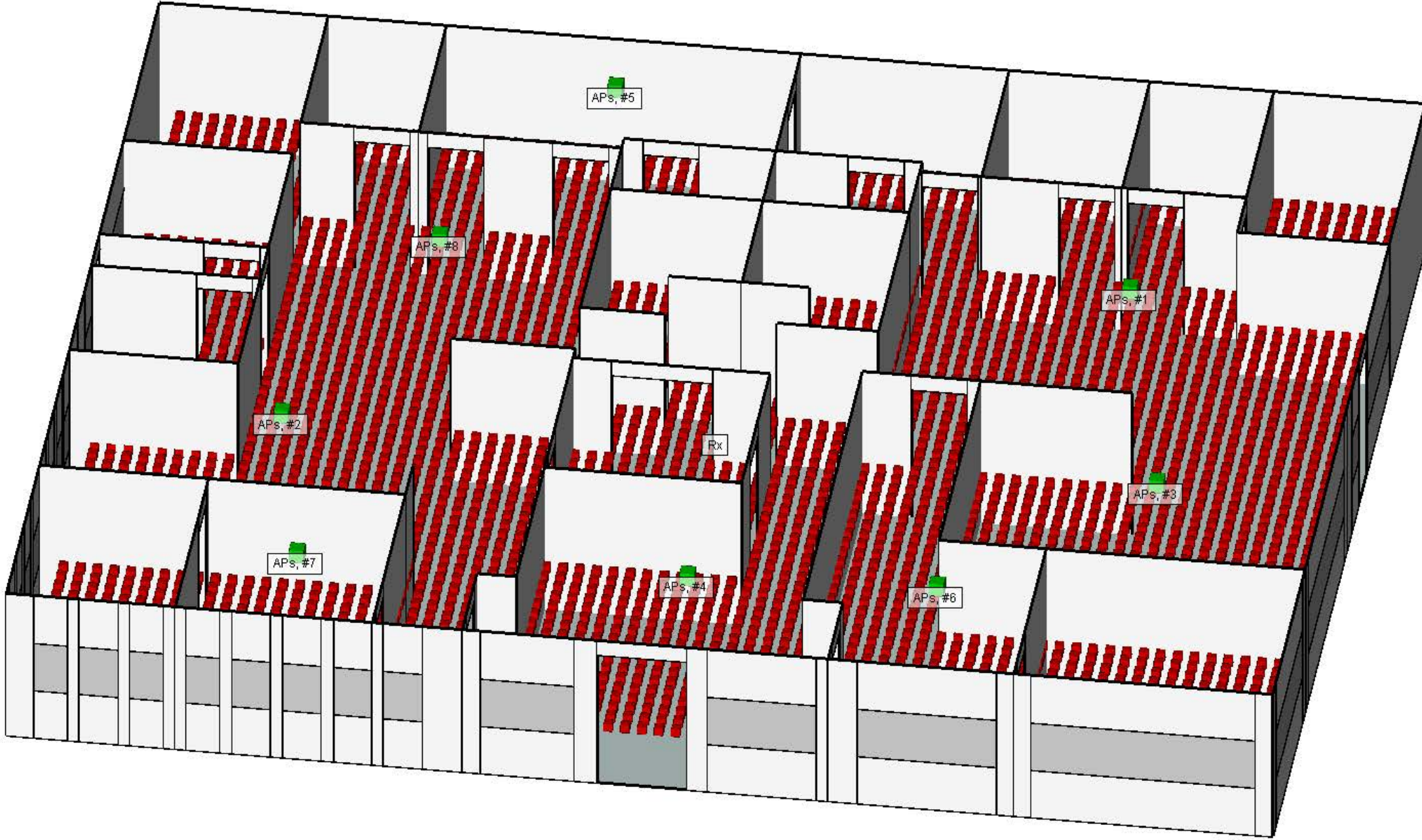}
    \label{fp2}
    \end{minipage}
    }
    \subfigure[]{
    \begin{minipage}[t]{0.35\textwidth}
    \centering
    \includegraphics[width=\textwidth]{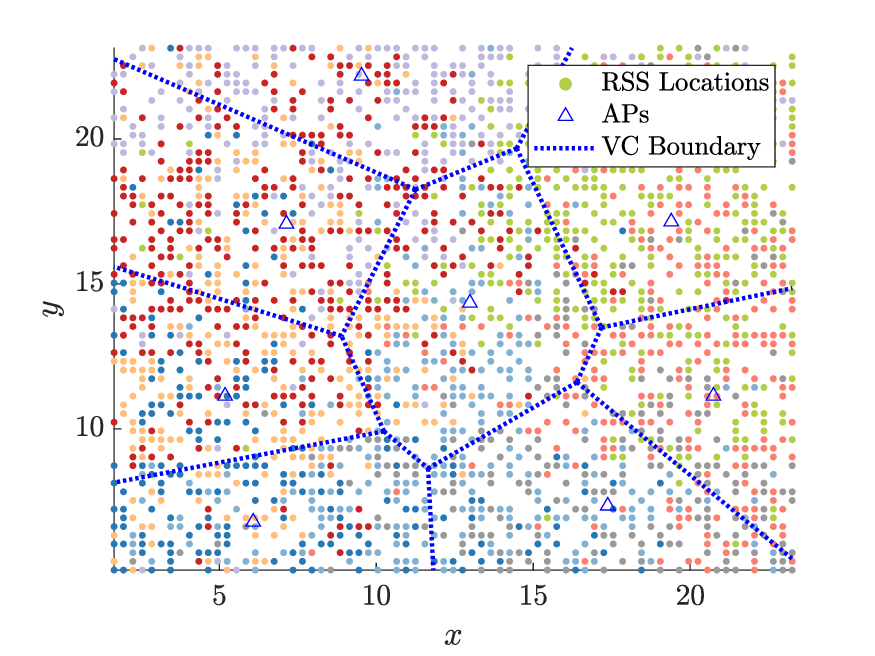}
    \label{cell_b}
    \end{minipage}
    }
	\caption{Illustrations of the three-dimensional ray-tracing setup using Wireless InSite. (a) Floor Plan 1: the basement of Agnew Hall; (c) Floor Plan 2: the first floor of Pack Building at Virginia Tech. The division into Voronoi cells and the ground-truth locations where the RSS measurements were obtained are shown in (b) for Floor Plan 1 and (d) for Floor Plan 2. The different colors of the RSS locations represent the clusters determined by the largest RSS value.}
    \label{fp}
\end{figure*}

\subsection{Experimental Settings}
We set up two indoor environments using wireless ray-tracing software Wireless InSite~\cite{Remcom2024}. These environments are built based on realistic floor plans at Virginia Tech's campus in Blacksburg, VA. 
Specifically,~\figref{fp1} illustrates Floor Plan 1, representing the basement of Agnew Hall with dimensions $31.1 \, \nrmm \times 19.6 \, \nrmm$, and~\figref{fp2} shows Floor Plan 2, representing the first floor of the Pack Building with dimensions $25 \, \nrmm \times 25 \, \nrmm$.
In our simulations, we assume that the windows are made of plain glass and the walls are constructed of concrete with a thickness of $0.3 \, \nrmm$. 
The ceiling is at a height of $3 \, \nrmm$ above the floor. 
We place $n = 8$ APs on the ceiling and $m \in \left\{4130, 4453\right\}$ potential receiver locations across the AOI at a height of $h \in \left\{1, 1.5\right\}\,\nrmm$ above the floor.
These measurement locations are distributed across different rooms on the same floor, as evident from~\figref{fp}.
The APs operate on channel 1 at $2.4\, {\rm GHz}$ and channel 149 at $5\, {\rm GHz}$, with bandwidths $B = 20 \, {\rm MHz}$, transmitting at a power of $30 \, {\rm dbm}$ using IEEE 802.11n protocols.
For each receiver location, the RSS values are measured for all APs.
The measurement noise is modeled using a log-normal distribution with a standard deviation of $3\, {\rm dB}$.
These configurations simulate complex indoor environments to evaluate the performance of our localization framework under realistic conditions.

\subsection{General Performance and Comparison}
\begin{figure*}
	\centering
	\subfigure[]{
    \begin{minipage}[t]{0.45\textwidth}
    \centering
    \includegraphics[width=\textwidth]{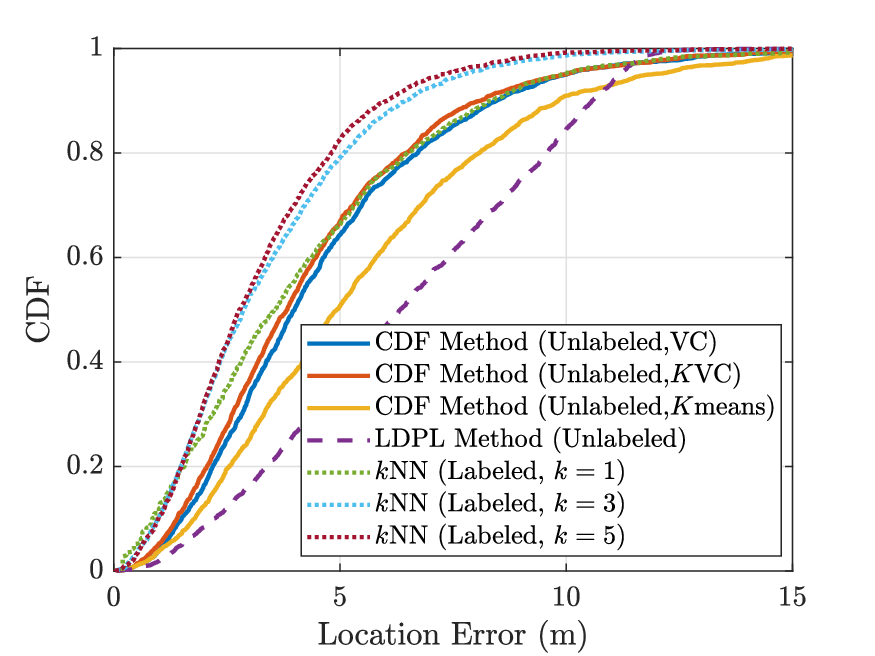}
    \label{cdf_comparison_p1}
    \end{minipage}
    }
    \subfigure[]{
    \begin{minipage}[t]{0.45\textwidth}
    \centering
    \includegraphics[width=\textwidth]{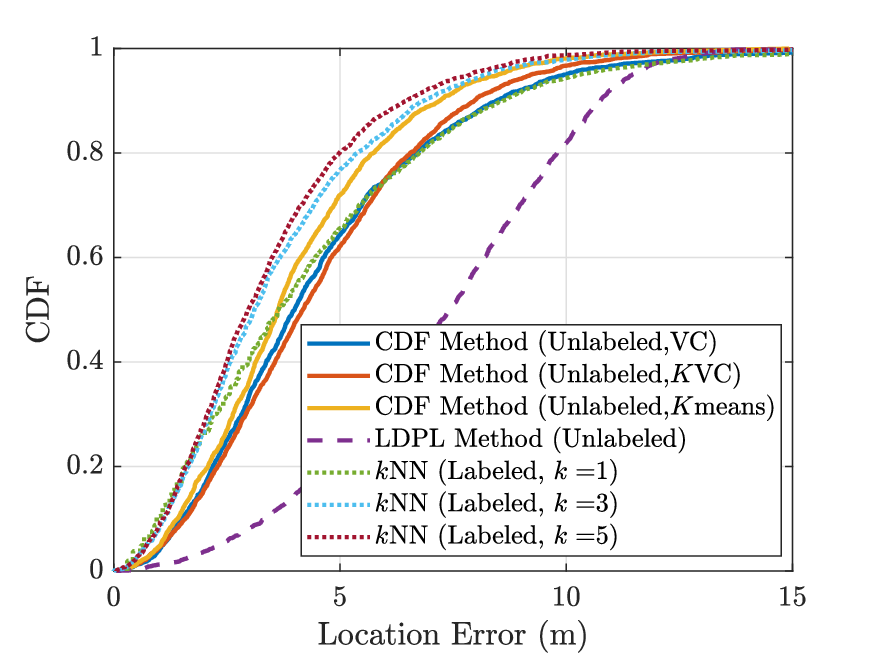}
    \label{cdf_comparison_p2}
    \end{minipage}
    }
	\caption{The comparison of the CDF of positioning errors using crowdsourced RSS dataset: (a) Floor Plan 1 with $m = 4130$ RSS samples and (b) Floor Plan 2 with $m = 4453$ RSS samples.}
\end{figure*}

We simulate signal propagation using the ray-tracing software described above and collect $m=4130$ crowdsourced data points for Floor Plan 1 and $m=4453$ for Floor Plan 2. Each dataset is split in half: one half provides statistical information, and the other half tests the localization performance.
We illustrate the locations of the APs and the measurement points, along with the division of Voronoi cells based on the AP locations, in~\figref{cell_a} and~\figref{cell_b}. The clusters of the crowdsourced RSS vectors are estimated using three clustering methods: Voronoi cells (VC), $K$ Voronoi cells ($K$VC), and $K$ means, respectively.
Generally, clustering errors occur due to wall blockages and signal reflections, which can affect the RSS values. However, even when some RSS vectors are incorrectly labeled, these incorrect labels typically correspond to nearby Voronoi cells. This proximity ensures that such incorrect clustering does not significantly impact the statistical learning of the path loss model using clustered RSS vectors, as these measurements are collected from nearby locations with similar environmental conditions.
We then perform the CDF conversion to derive the distances to each AP based on the RSS measurements in each cluster.
The locations of the measurements are estimated using the converted distances through trilateration.
For comparison, we employ two baseline methods. 
The first method is an RSS conversion approach based on the LDPL model~\cite{koo2012unsupervised}, which utilizes crowdsourced RSS data without location annotations. This method linearly converts RSS to distance by mapping the highest RSS to distance zero and the lowest RSS to a known maximum coverage distance.
The second is the $k$NN algorithm~\cite{7883033}, which uses the same RSS measurements but with ground truth location labels. These baselines are used to estimate distances and evaluate localization performance against our proposed method.
The CDFs of localization errors are plotted in~\figref{cdf_comparison_p1} and~\figref{cdf_comparison_p2}. 
The CDF curves demonstrate that our CDF conversion method outperforms the linear conversion method, providing more accurate localization for indoor measurements.
Notably, our method's localization accuracy approaches that of the $k$NN algorithm using fingerprints with ground-truth location labels.
The result highlights the effectiveness of the CDF conversion method in capturing signal propagation characteristics and provides reasonably high accuracy.

\subsection{Generalization Performance and Comparison}
We further evaluate the generalization performance of our localization framework on Floor Plan 1 under two different settings:

\subsubsection{Frequency Generalization}

\begin{figure}
	\centering
    \includegraphics[width=0.45\textwidth]{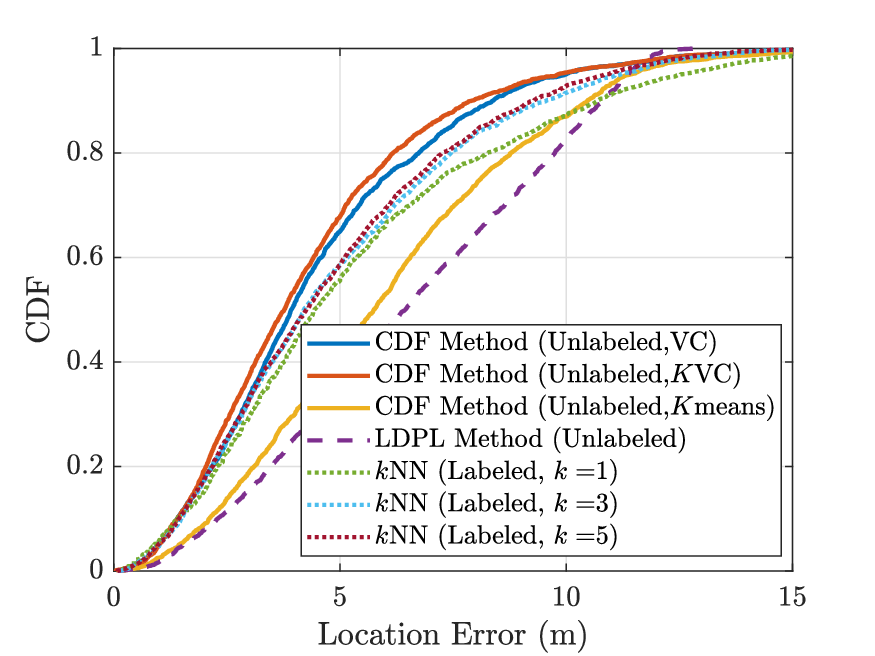}
	\caption{The comparison of the CDF of positioning errors using crowdsourced RSS dataset across different operating frequencies, with $m = 2065$ RSS samples at $2.412\,{\rm GHz}$ and $5.745\, {\rm GHz}$, respectively.}
    \label{p3}
\end{figure}

We aim to evaluate the ability of our method to generalize across different operating frequencies where signal propagation characteristics differ. This evaluation tests the robustness of our framework in adapting to frequency variations without requiring retraining or collecting new data at the target frequency. 
We collect $m = 2065$ crowdsourced RSS measurements at $2.412\,{\rm GHz}$ and evaluate the localization performance based on $m = 2065$ RSS measurements obtained at $5.745\, {\rm GHz}$.
The results are depicted in~\figref{p3}, showing the CDF of positioning errors for our proposed method and baseline methods. 
Our method demonstrates strong generalization performance across frequencies, achieving the lowest localization error compared to the baseline approaches. 
This performance indicates that the statistical learning approach in our framework effectively captures the underlying propagation characteristics, enabling accurate localization even when transitioning to a different frequency band.
In contrast, the $k$NN algorithm shows a noticeable degradation in performance, as it is highly sensitive to the specific features of the frequency used during data collection. 
These findings highlight the robustness of our method in adapting to variations in operating frequencies, demonstrating its potential for real-world applications where cross-frequency localization is required.

\subsubsection{Height Generalization}
\begin{figure}
	\centering
    \includegraphics[width=0.45\textwidth]{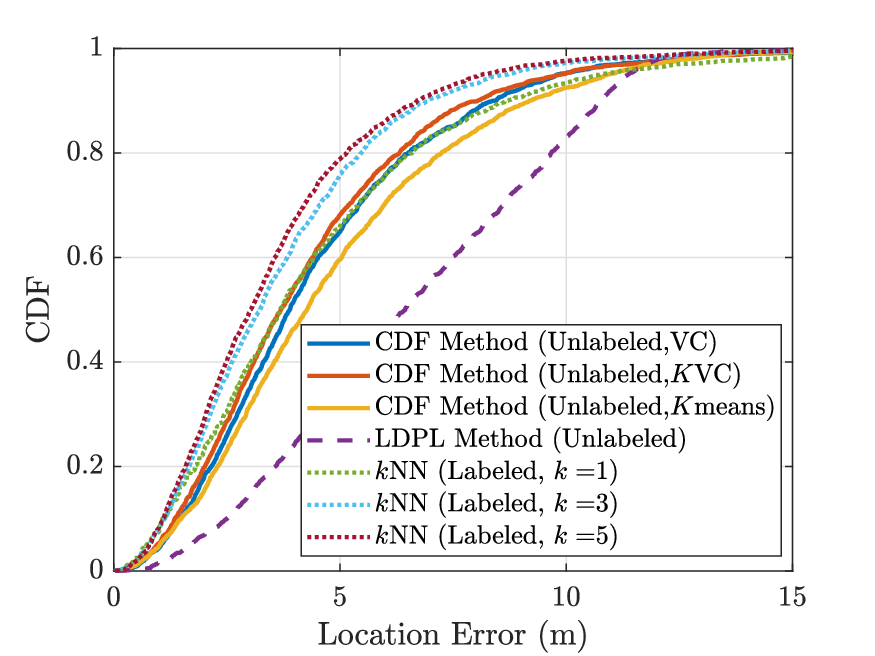}
	\caption{The comparison of the CDF of positioning errors using crowdsourced RSS dataset across different receiver heights, with $m = 2065$ RSS samples at a height of $1\,\nrmm$ and $1.5\,\nrmm$, respectively.}
    \label{p4}
\end{figure}

We further examine the generalization performance of our method to different receiver heights. The crowdsourced data is collected at the height of $1\,\nrmm$, representing a typical handheld device in a pocket. At the same time, the localization is then performed at the height of $1.5\,\nrmm$, simulating scenarios where devices are held at shoulder level or mounted at higher positions. 
This test evaluates the effectiveness of our framework in handling vertical variations in the environment.
Our method demonstrates strong generalization performance when transitioning from a receiver height of $1\,\nrmm$ to $1.5\,\nrmm$. 
Despite the differences in signal propagation characteristics due to the change in height, the CDF of positioning errors indicates that our localization framework maintains high accuracy.

\subsection{Performance with Side Information}

\begin{figure}
	\centering
    \includegraphics[width=0.45\textwidth]{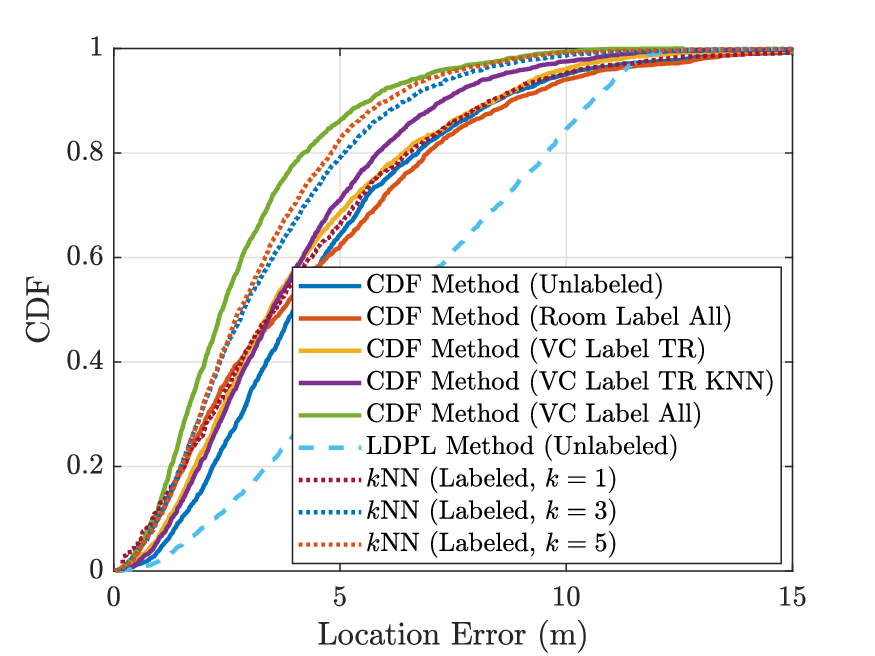}
	\caption{The comparison of the CDF of positioning errors using crowdsourced RSS dataset with side information.}
    \label{p5}
\end{figure}

In this section, we evaluate the performance of our localization framework when additional side information is available alongside the crowdsourced RSS data. Specifically, we consider scenarios where the collected RSS data includes partial labeling, such as the room in which the RSS measurements were collected or the corresponding Voronoi cell for each measurement.
We analyze two cases: (1) partial labels are available only for the training data, and (2) partial labels are available for both the training and testing data.
This label information is used to cluster the RSS measurements, and the CDF-based distance estimation is applied separately to each cluster.
In the first case, we evaluate the model's ability to learn the ground truth accurately from labeled training data and apply the learned model to estimate locations for crowdsourced RSS measurements without auxiliary labels.
In the second scenario, all measurements are correctly assigned to their corresponding rooms or Voronoi cells, and the capability of our distance estimation algorithm is evaluated.
The localization performances with and without side information are presented in~\figref{p5}.
The inclusion of side information enhances the performance of our localization framework. 
Notably, when Voronoi cell labels are available for both the training and testing data, the localization performance exceeds that of the $k$NN algorithm utilizing location labels. 
However, with room labels, the improvement is limited to achieving lower distance errors rather than overall performance gains.
When labels are available only for the training data, improvements in localization error are observed, particularly when combined with the $k$NN algorithm. In this approach, clusters of crowdsourced RSS measurements are estimated, and the models of top $K$ matching clusters are used to estimate distances, which are then averaged to determine the location.
These results present the potential of hybrid approaches that integrate crowdsourced data with varying levels of side information to improve localization performance.

\section{Conclusion}
This paper introduces a novel calibration-free framework for indoor positioning that utilizes unlabeled crowdsourced RSS data. The framework consists of clustering RSS vectors with Voronoi cells, converting the RSS values into distance estimates, and determining the target's location using an improved trilateration method.
Our results show that the localization accuracy of our proposed framework without location labels approaches that of the $k$NN algorithm, which relies on fingerprints with location labels. Our framework is highly suitable for real-world applications, as it removes the need for labor-intensive site surveys and scales effectively to large unlabeled datasets. 
Moreover, the framework demonstrates robustness to variations in operating conditions, such as different frequencies and receiver heights, highlighting its adaptability to diverse indoor environments.

Future work could explore advanced models for RSS propagation, incorporate dynamic environmental changes, and combine the framework with other range-based localization methods. 
Additionally, hardware implementations could validate its practical use, bridging the gap between theory and real-world deployment. This work provides a strong foundation for efficient, scalable, and accurate indoor localization systems.

\appendices
\section{Order Statistics} \label{app-c2}
Assume that the RSS is a monotonically decreasing function of distance from AP within a given interval $[a, b]$.
We estimate the locations where the measurements  $q_1, q_2, \cdots, q_n$ were obtained.
These measurements are assumed to be a monotonically decreasing function of distances $q = g(d)$, where the distances  $d_1, d_2, \cdots, d_n$ are uniformly sampled on the interval $[a, b]$. 
We first sort the measurements in descending order, such that 
\begin{align*}
    q_{(1)} \geq q_{(2)} \geq \cdots \geq q_{(n)}.
\end{align*}
Due to the monotonic assumption, the largest RSS values correspond to the smallest distances and vice versa. Based on this ordering, we estimate the distances by mapping the ordered RSS values to evenly spaced quantiles of the uniform distribution over $[a, b]$. 
The estimated distances are computed as:
\begin{align}
    \hat{d}_{(r)} = a + (b - a) \frac{r}{n+1},
\end{align}
where $\hat{d}_{(r)}$ is the distance estimation for the $r$-th largest RSS measurements.
To prove that $\hat{d}_{(r)}$ is an unbiased estimator of $d_{(r)}$, we analyze the statistical properties of $d_{(r)}$. The probability density function (PDF) of the $r$-th order statistic $D_{(r)}$ is
\begin{align}
    f_{D_{(r)}}(x) = \frac{n!}{(r-1)! (n-r)!} &f_D(x) \\
    \cdot [F_D(x)]^{r-1}& [1 - F_D(x)]^{n-r},
\end{align}
where the uniform distribution over $[a,b]$ is characterized by
\begin{align}
    f_D(x) = \frac{1}{b-a}\delta(a<x<b), \\
    F_D(x) = \frac{x-a}{b-a}\delta(a<x<b). 
\end{align}
Substituting these into the order statistic PDF, we obtain:
\begin{align}
    f_{D_{(r)}}(x) = \frac{(b-a)^{-n} n!}{(r-1)! (n-r)!} [x-a]^{r-1} [b-x]^{n-r}.
\end{align}
Defining the transformation $y_{(r)} = \frac{d_{(r)}-a}{b-a}$, we recognize that $Y_{(r)}$ follows a Beta distribution with parameters $\alpha = r$ and $\beta = n-r-1$.
Using the properties of the Beta distribution, we find:
\begin{align}
    \nbbE [D_{(r)}] &= \nbbE [a + (b-a)Y_{(r)}] \\
    &= a + (b-a) \frac{\alpha}{\alpha + \beta} \\
    &= a + (b-a) \frac{r}{n+1}.
\end{align}
This confirms that $\hat{d}_{(r)}$ is an unbiased estimator of $d_{(r)}$.
Similarly, we can derive the variance, given by: 
\begin{align}
    \nbbV [D_{(r)}] &= \nbbV [a + (b-a)Y_{(r)}] \\
    &= (b-a)^2 \frac{\alpha\beta}{(\alpha+\beta)^2 (\alpha+\beta+1)} \\
    &= (b-a)^2 \frac{r(n-r+1)}{(n+1)^2(n+2)}.
\end{align}
This shows that the mean squared error (MSE) depends on $r$, implying that different order statistics yield different estimation accuracies. In particular, the estimation error is largest when $r$ approaches $n/2$ and smallest when $r$ is close to $1$ or $n$.


\bibliographystyle{IEEEtran}
\bibliography{citation}

\end{document}